\documentclass[aps,pra,twocolumn, 10pt,longbibliography, floatfix]{revtex4-2}

\usepackage[utf8]{inputenc}
\usepackage[caption=false]{subfig}

\usepackage{verbatim}
\usepackage{booktabs}
\usepackage{adjustbox}
\usepackage[version=4]{mhchem}
\usepackage{booktabs}

\usepackage{listings}

\usepackage{ragged2e} 

\usepackage[table,xcdraw]{xcolor} 

\usepackage{enumitem}
\usepackage[nopatch]{microtype}
\usepackage{balance}

\usepackage{algorithm}
\usepackage{algpseudocode}

\usepackage{makecell}
\usepackage{mhchem}

\usepackage{soul}
\usepackage{url}
\usepackage{mathtools}
\usepackage{braket}

\usepackage{dsfont}
\usepackage[english]{babel}
\usepackage{graphicx} 
\usepackage{tikz}
\usetikzlibrary{quantikz2}

\usepackage{hyperref}

\usepackage{natbib} 
\usepackage{physics}

\usepackage{listings}

\usepackage{amssymb}

\usepackage{babel}

\newcommand{\LDIG}{
Lighthouse Disruptive Innovation Group, LLC
1 Broadway, 14th floor, Cambridge,
Middlesex County, Massachusetts 02142 (USA)
}

\newcommand{\LDIGEU}{
Lighthouse Disruptive Innovation Group Europe, SL.
Barcelona - Spain
}

\newcommand{\MIT}{
MIT Media Lab - City Science Group, Cambridge, USA
}

\renewcommand{\arraystretch}{2} 

\graphicspath{{./images/}}

\begin{document}
\title{QMProt: A Comprehensive Dataset of Quantum Properties for Proteins}

\author{Laia Coronas Sala}
\affiliation{\LDIGEU}
\email{laia.coronas@lighthouse-dig.com}

\author{Parfait Atchade-Adelomou}
\affiliation{\LDIGEU}
\affiliation{\LDIG}
\email{parfait.atchade@lighthouse-dig.com}
\affiliation{\MIT}
\email{parfait@mit.edu}

\date{April 2025}

\begin{abstract}

\noindent We introduce Quantum Mechanics for Proteins (QMProt), a dataset developed to support quantum computing applications in protein research. QMProt contains precise quantum-mechanical and physicochemical data, enabling accurate characterization of biomolecules and supporting advanced computational methods like molecular fragmentation and reassembly. The dataset includes 45 molecules covering all 20 essential human amino acids and their core structural elements: amino terminal groups, carboxyl terminal groups, alpha carbons, and unique side chains. QMProt primarily features organic molecules with up to 15 non-hydrogen atoms (C, N, O, S), offering comprehensive molecular Hamiltonians, ground state energies, and detailed physicochemical properties. Publicly accessible, QMProt aims to enhance reproducibility and advance quantum-enhanced simulations in molecular biology, biochemistry, and drug discovery.


\end{abstract}

\maketitle

\section{INTRODUCTION}

Quantum mechanics (QM) plays a crucial role in the accurate modeling of biomolecules. By providing insights into their structure, functions, and interactions, QM enhances our understanding of complex systems such as proteins, potentially improving current drug discovery processes \cite{baiardi2023quantum, vanmourik2004firstprinciples, usman2021peptides}. However, proteins—one of the most structurally diverse and functionally significant classes of biomolecules—pose considerable challenges due to their size and complexity, requiring a large number of qubits for accurate simulation \cite{nielsen2010quantum, atchadeadelomou2022quantum}. To overcome these challenges, several strategies have been proposed.

One promising approach is fragmentation. In the case of peptides, fragmentation primarily involves the simulation of individual amino acids followed by their reassembly, while accounting for interactions and applying chemical corrections \cite{collins2006, li2005, deev2005, bettens2006}. Building upon this research direction, our previous work introduced a disruptive strategy for fragmenting peptides into computationally feasible amino acids and reassembling them post-simulation, incorporating chemical corrections related to bond formation \cite{CoronasSala2025}, obtaining very promising results.

On the other hand, there has been a rapid advancement in Artificial Intelligence (AI), Machine Learning (ML), and Quantum Machine Learning (QML), with these emerging as promising approaches for predicting quantum properties in larger systems \cite{batra2021quantum, tkatchenko2020machine}. Acknowledging this, several large datasets have been developed to train such algorithms, typically including a vast number of molecules and isomers.

One of the largest is QM7-X, a dataset comprising $4.2$ million small organic molecules with up to seven non-hydrogen atoms \cite{hoja2021}, which has proven highly useful for predicting ground-state properties \cite{coronas2024leveraging}. Prior to its development, the QM8 and QM9 datasets also provided extensive coverage of quantum properties across a large set of molecules \cite{ramakrishnan2014, ramakrishnan2015, ruddigkeit2012}. QMugs is another specialized dataset, specifically designed for ML-driven studies on drug-like molecules \cite{issert2022}. Additional datasets including various small molecules have also been introduced to further advance research in this area, as summarized in this review \cite{ullah2024molecular}. 

Analyzing the state of the art, we identified a significant gap in datasets containing larger molecules, particularly organic compounds such as proteins, which play a crucial role in numerous biological processes \cite{nelson2021lehninger, alberts2015molecular}. Existing datasets primarily focus on small molecules, making it extremely challenging to extrapolate properties to much larger biomolecular systems \cite{coronas2024leveraging}. Moreover, while these datasets are undoubtedly valuable for ML and QML applications, they do not provide solutions to harness QM in proteins. Therefore, the motivation behind QMProt is to bridge this gap by providing a robust and efficient dataset of molecules and features, potentially leading to the following contributions:

\begin{itemize}
\item Facilitating research on relevant organic molecules by including crucial yet computationally expensive properties, such as ground state energy and the molecular Hamiltonian, accelerating and fostering advancements in quantum simulations of biomolecules.

\item Enhancing the characterization of larger biomolecular systems by bridging the gap between existing datasets—primarily focused on small molecules—and the needs of researchers working on larger systems, such as peptides and proteins.

\item By providing a dataset that integrates QM-derived properties with ML methodologies, QMProt enables hybrid QM/ML approaches, enabling researchers to train models that accurately and efficiently predict the properties of larger and more complex systems.

\item Accelerating drug discovery and biomolecular research, as proteins are central to numerous biological and therapeutic processes.

\item Enabling the study of fragmentation and reassembly techniques, proposing new chemical corrections for bond formation and ensuring the accurate reconstruction of molecular properties post-simulation, aligning with the results obtained in our latest work \cite{CoronasSala2025}.
\end{itemize}

\section{Methodology} \label{sec:methodology}

\begin{figure}[ht]
    \centering
    \includegraphics[width=1\linewidth]{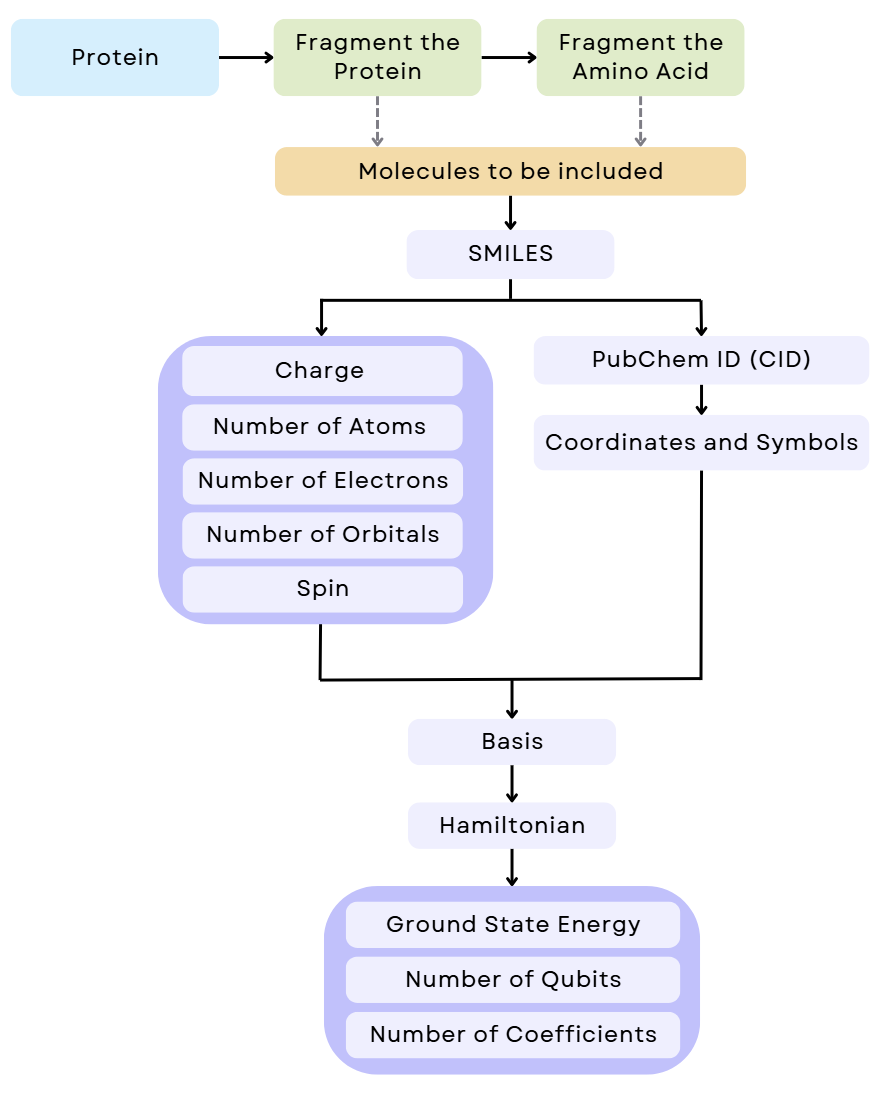}
    \caption{\justifying General pipeline followed for the molecular inclusion and property computation to form this dataset. Starting from the top: the included molecules were those obtained from the fragmentation of any protein or amino acid, thus covering any possible peptide. For each molecule, the names were stored, as well as the formal abbreviation in the case of amino acids. Then, using the SMILES string of each molecule, the CID, symbols, and coordinates were directly obtained from PubChem \cite{pubchem}, and properties such as charge, number of atoms, electrons, orbitals, and spin were computed. Lastly, a specific basis was chosen to represent each molecule and calculate its Hamiltonian operator, number of qubits, number of coefficients, and ground state energy.}
    \label{fig:1}
\end{figure}

Figure \ref{fig:1} illustrates the general pipeline used for molecular inclusion into the dataset, as well as the process followed to obtain each of the features.

\subsection{Molecular Inclusion Criteria}

The aim of the molecules included in the dataset was to cover as many possible fragments resulting from protein fragmentation, in order to facilitate the understanding and study of these molecules. Consequently, the chosen molecules were those generated from the fragmentation of proteins and amino acids.The upper part of Figure \ref{fig:1} perfectly represents this process.

First, since all proteins can be fragmented into the same 20 amino acids, these molecules were included. Then, given that a single amino acid can be computationally demanding, we further fragmented each amino acid into three main groups: the amino terminal, the carboxyl terminal, and the alpha carbon, as well as their corresponding 20 different side chains. All of these molecules were included in the dataset.

Additionally, we incorporated small molecules such as \ce{H2O}, \ce{H2}, and \ce{CH3}, since they are common molecules involved in group addition and bond formation.

In total, $45$ molecules were included in the dataset. While more molecules could be generated, this selection was made to minimize errors that could arise from further fragmentation and to provide a straightforward and focused dataset for protein fragmentation, rather than an overly extensive one.

\subsection{Properties Included in the Dataset}

For each included molecule, we provide a series of descriptive, physicochemical, and quantum properties for accurate characterization. Below is a short description of each variable included in the dataset.

\begin{itemize}
    \item Abbreviation: this is only present for molecules corresponding to entire amino acids, as it is a formalism used to refer to these molecules. For instance, \textit{Histidine} is commonly referred to as \textit{His}.
    
    \item Name: this corresponds to the complete common name of the molecule. 
    
    \item Molecular formula (mf): this corresponds to the compact SMILES string of the molecule. It is a formalism for grouping and counting the atoms that make up the molecule. 
    
    \item CID: unique identifier of the molecule in the PubChem database \cite{pubchem}. This is found by inputting the SMILES string in the search bar in PubChem, and selecting the CID corresponding to the correct conformation of the molecule of interest. 
    \item Number of atoms: this is directly computed by adding all the elements in the SMILES string. 
    
    \item Charge: the charge is a direct consequence of the amino acids forming the molecule. In most cases, we considered the molecules to be in a neutral state, even though under certain conditions they might be prone to ionization.
    
    \item Number of electrons: this property was directly computed from the SMILES string. The number of electrons was considered as follows: $9$ for carbon, $1$ for hydrogen, $8$ for oxygen, $7$ for nitrogen, and $16$ for sulfur, providing with an idea of the complexity of the quantum properties to be solved.

    \item Number of orbitals: This is directly related to the energy levels of the molecule and the distribution of its electrons.
    
    \item Bond length: the bond length is defined as the minimum value of the distance matrix \cite{ccbdb}, which is calculated based on the $3$D positions of the atoms. The Euclidean distance formula calculates the distance between each pair of atoms. Therefore, the bond length is computed as shown in Equation \ref{eq:bondlength}.

\begin{equation}
    \text{Bond length} = \min \left( \{ d_{ij} \mid i \neq j \} \right)
    \label{eq:bondlength}
\end{equation}

    Where $d_{ij}$ is the distance with Euclidean norm between atoms \( i \) and \( j \). 
  
    \item Coordinates: the $3$D coordinates of all atoms in the molecule were extracted directly from PubChem as SDF files and reorganized into the h5 files.
    
    
    \item Spin: the spin of the molecule is directly obtained from the SMILES string by determining the number of unpaired electrons according to the atoms forming the molecule. In general, the spin was considered $0$ for most entire amino because it was assumed all the electrons were paired, while for several radical molecules, the spin was $1$.
    
    \item Basis: for simplicity, and since most molecules involved considerable computational complexity, we employed the STO-3G basis representation for most molecules.

    \item Number of qubits: number of qubits required for the quantum simulation of the molecule. This depends on the encoding scheme and the complexity of the molecular system, therefore determining the computational resources needed for quantum calculations.

    \item Number of coefficients: this represents the total number of terms in the molecular wave function expansion. Typically, a higher number of coefficients generally leads to more accurate representations but also increases computational cost.

    \item Hamiltonian: the Hamiltonian of the molecule was computed using the coordinates, charge, spin, and basis set. The Hamiltonian is crucial for molecular characterization, as it provides insights into the energetic state and time evolution of the system. However, its computation can be challenging and time-consuming for larger molecules, therefore, we decided to directly provide it.
   
    \item Energy: the energy refers to the ground state energy of the molecule in Hartrees, corresponding to its most stable configuration. This offers insights into molecular stability and potential interactions. Furthermore, similar to the Hamiltonian, this property is time-consuming to compute, so we provide it directly to facilitate further studies on protein and biomolecular characterization.
\end{itemize}

\subsection{Validation}

As mentioned, most properties were directly computed from sources from the literature \cite{pubchem, ccbdb} or the SMILES string of the molecules. However, others required more complex methods, such as the computation of the Hamiltonian and ground state energies.

Hamiltonian calculations were performed using the OpenFermion library \cite{McClean2020}, entering the basis set (STO-3G in most cases), charge, and multiplicity, which were calculated from the given spin ($M = 2S + 1$), along with the molecular coordinates. The molecule was then processed using PySCF and self-consistent field (SCF) theory \cite{pyscf}, and the final computed Hamiltonian was transformed into fermionic format (according to Pennylane standards) \cite{Bergholm2018}. 

Energy calculations were performed using the well-established Hartree-Fock (HF) methodology, leveraging the precision achievable with today’s classical computing capabilities to ensure a robust baseline, independent of potential advancements in future quantum computing \cite{pyscf_scf}. The $3$D atomic coordinates and molecular system types were extracted from PubChem SDF files \cite{pubchem} and processed using the PySCF package for HF calculations \cite{pyscf_scf}. Specifically, we employed the restricted Hartree-Fock (RHF) method with a minimal basis set (STO-3G in most cases) to compute the total ground state energy of the system.

Energy calculations were performed on a high-performance computing environment to ensure quantum simulations' efficiency and precision. The primary computational setup consisted of a 13th Gen Intel\textsuperscript{\textregistered} Core\textsuperscript{TM} i7-13700H processor with 32 GB of RAM, running a 64-bit operating system under the Windows Subsystem for Linux (WSL). This configuration facilitated initial processing tasks and preliminary computations.

A dedicated high-performance server was employed for more intensive calculations, particularly those involving Hamiltonian operators. This system featured a 24-core AMD Threadripper Pro 5965WX processor operating at 3.80 GHz, providing substantial parallel processing capabilities. Additionally, three NVIDIA RTX 6000 Ada GPUs, each equipped with 48 GB of VRAM, were utilized to accelerate matrix operations and tensor contractions essential for quantum state evolution. To support the high memory demands of these calculations, the system was equipped with 256 GB of RAM distributed across two 128 GB 3200 MHz DDR4 ECC/REG modules, ensuring both speed and reliability in handling large-scale quantum data.

This computational infrastructure provided the necessary resources for efficient quantum simulations, allowing for precise energy calculations and the manipulation of complex Hamiltonian matrices in a high-dimensional space.

\section{Data Records}


The QMProt dataset is provided as 45 different H5 files, each containing molecular properties as attributes. A README file is also included that provides technical details on accessing the information. Within the H5 files, each attribute represents a different molecular property previously described. We have also organized the attributes hierarchically so that the Molecule attribute includes the properties: symbols, coordinates, charge, basis, and spin, formatted for PennyLane.  

Additionally, due to the size of some molecular Hamiltonians, certain Hamiltonians had to be partitioned into multiple attributes named $hamiltonian\_1$, $hamiltonian\_2$, and so on. To obtain a full representation of the system, these attributes should be concatenated in the correct order. In our GitHub repository, we provide the code for this concatenation as well as for converting the Hamiltonians into a PennyLane operator \cite{pennylane}.  

\begin{figure}[!ht]
    \centering
    \includegraphics[width=1\linewidth]{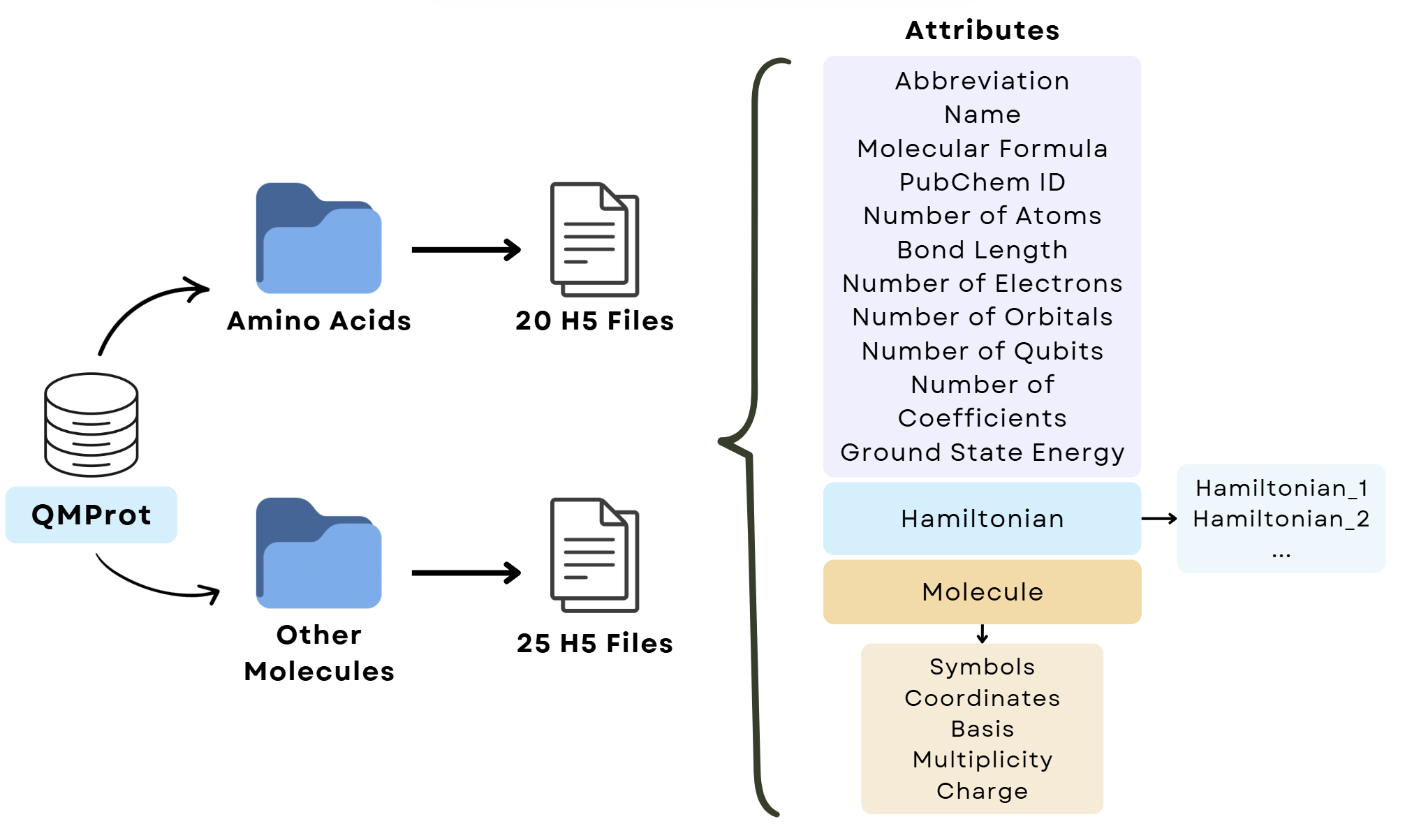}
    \caption{\justifying Structure of the QMProt dataset. QMProt comprises 45 different h5 files that include all the attributes corresponding to the molecular properties described.}
    \label{fig:2}
\end{figure}

This format allows for efficient querying and manipulation, facilitating the application of models and statistical studies on molecular properties. Figure \ref{fig:2} illustrates the structure of the dataset, showing groups and attributes for each molecule.

\begin{figure}[!ht]
    \centering
    \includegraphics[width=0.7\linewidth]{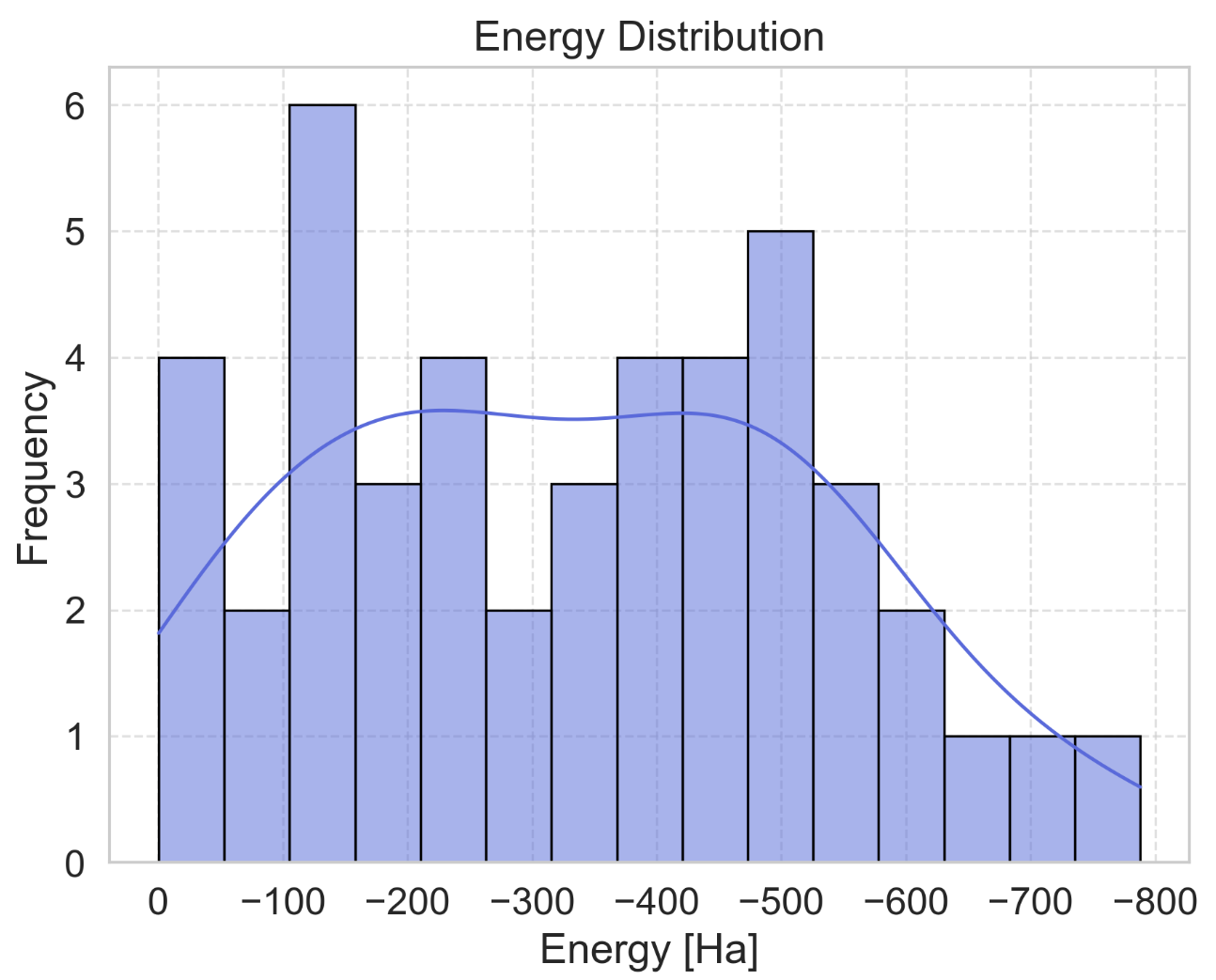}
    \caption{Distribution of the molecular energies included in the dataset.}
    \label{fig:3}
\end{figure}

\begin{figure}[!ht]
    \centering
    \includegraphics[width=0.8\linewidth]{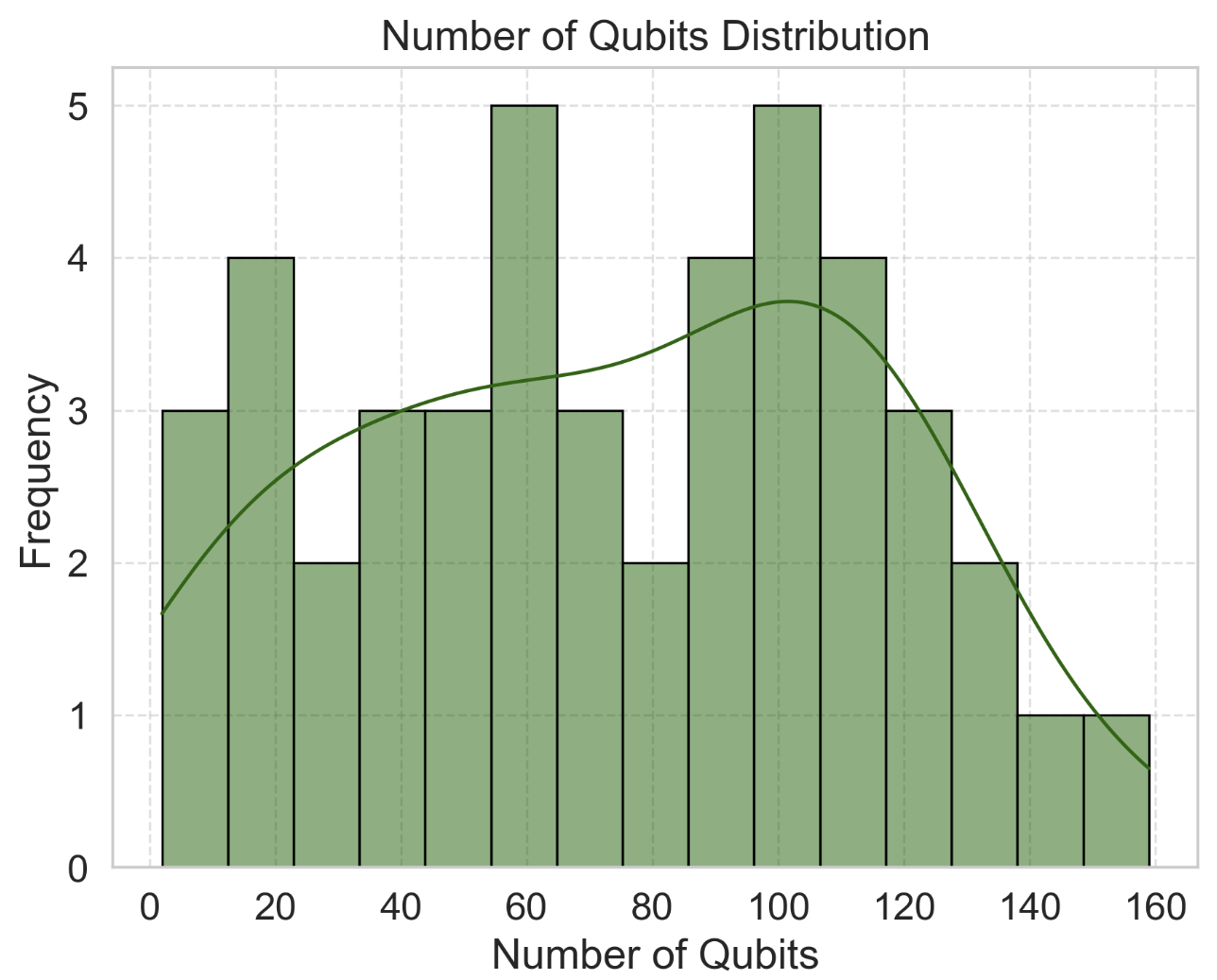}
    \caption{Distribution of the number of atoms of the molecules included in the dataset.}
    \label{fig:4}
\end{figure}

Furthermore, to have a broader view of the molecules included in the dataset, Figures \ref{fig:3} and \ref{fig:4} show the distribution of the energies and the number of atoms of the included molecules, respectively.

Ultimately, the data corresponding to the final included molecules and their size in terms of electrons, orbitals, qubits, and coefficients can be seen in Table \ref{tab:molecules}.

\section{Conclusions}\label{sec:conclusions}

The QMProt dataset represents a significant step forward in the quantum simulation of biomolecules, particularly proteins. By providing a comprehensive collection of $45$ carefully selected molecules, including the $20$ essential amino acids and their relevant subgroups, QMProt bridges a critical gap in current quantum chemistry datasets.

This dataset is designed to enhance the accuracy and efficiency of quantum simulations for larger biomolecular systems. It includes detailed molecular properties such as bond lengths, atomic coordinates, spin states, and the calculated Hamiltonian, offering an extensive foundation for researchers working in protein folding, drug discovery, and biomolecular interactions. Unlike previous databases that primarily focus on providing large-scale datasets for Machine Learning applications, QMProt emphasizes the precise molecular characterization of important molecules, particularly amino acids, laying a solid foundation for future advancements in the study of larger molecules.

Lastly, through this work, we present a valuable tool for quantum simulations and propose an innovative approach to the fragmentation and reassembly of proteins, enabling the accurate prediction of quantum properties in large, complex biomolecules. This approach builds on our previous work, where we proposed a strategy for reassembling amino acids by applying chemical corrections to reconstitute protein properties that are otherwise difficult to compute. This is of the utmost importance, since the integration of chemical corrections with advanced quantum computational methods provides a basis for future advancements in protein simulations and related fields as we await the development of more powerful quantum computers.

In conclusion, QMProt will undoubtedly serve as a vital resource, promoting advancements in computational biology and quantum computing applications in biomolecular research. We hope that QMProt will inspire further efforts to develop comprehensive datasets that enable the integration of quantum mechanics in the study of larger and more complex biomolecular systems, such as proteins.

\vspace{7mm}

\section*{Code and Data Availability}
All source code, analysis scripts, and the complete QMProt dataset are openly available on GitHub at:
\newline
\url{https://github.com/LDIG-US/qmprot}
\newline
This repository includes tools for data preprocessing, visualization, and reproducibility of the experiments presented in this work.
\newline
The QMProt dataset is also hosted on the Pennylane platform for direct use in quantum pipelines at:
\newline
\url{https://pennylane.ai/datasets/collection/qmprot}
\newline

\section*{Acknowledgements} 
The authors gratefully acknowledge Guillermo Alonso-Linaje and Diego Guala for their valuable insights and contributions throughout the experimental process and dataset development. Special thanks are also extended to the Pennylane team for their technical support and thoughtful discussions, which significantly informed the design and implementation of the QMProt dataset.

\bibliographystyle{unsrt}
\bibliography{references}

\begin{table*}[htbp]
    \centering
    \renewcommand{\arraystretch}{1.2}
    \captionsetup{skip=10pt}
    \caption{Properties of different molecules and functional groups.}
    \label{tab:molecules}
    \begin{tabular}{|l|l|r|r|r|r|}
        \hline
        \textbf{Name} & \textbf{Formula} & \textbf{Electrons} & \textbf{Orbitals} & \textbf{Qubits} & \textbf{Coefficients} \\ \hline
        Histidine & C$_6$H$_9$N$_3$O$_2$ & 82 & 64 & 128 & 23831261 \\ \hline
        Leucine & C$_6$H$_{13}$NO$_2$ & 72 & 58 & 116 & 16200242 \\ \hline
        Isoleucine & C$_6$H$_{13}$NO$_2$ & 72 & 58 & 116 & 16379995 \\ \hline
        Lysine & C$_6$H$_{14}$N$_2$O$_2$ & 80 & 64 & 128 & 23906497 \\ \hline
        Methionine & C$_5$H$_{11}$NO$_2$S & 80 & 60 & 120 & 17802421 \\ \hline
        Phenylalanine & C$_9$H$_{11}$NO$_2$ & 88 & 71 & 142 & 36125918 \\ \hline
        Threonine & C$_4$H$_9$NO$_3$ & 64 & 49 & 94 & 8355908 \\ \hline
        Tryptophan & C$_{11}$H$_{12}$N$_2$O$_2$ & 108 & 87 & 159 & 92412988 \\ \hline
        Valine & C$_5$H$_{11}$NO$_2$ & 64 & 51 & 102 & 9819598 \\ \hline
        Arginine & C$_6$H$_{14}$N$_4$O$_2$ & 94 & 74 & 114 & 41609123 \\ \hline
        Cysteine & C$_3$H$_7$NO$_2$S & 66 & 46 & 92 & 6193299 \\ \hline
        Glutamine & C$_5$H$_{10}$N$_2$O$_3$ & 78 & 60 & 120 & 18268397 \\ \hline
        Asparagine & C$_4$H$_8$N$_2$O$_3$ & 70 & 57 & 106 & 11309980 \\ \hline
        Tyrosine & C$_9$H$_{11}$NO$_3$ & 96 & 76 & 102 & 46746137 \\ \hline
        Serine & C$_3$H$_7$NO$_3$ & 56 & 42 & 84 & 4532699 \\ \hline
        Glycine & C$_2$H$_5$NO$_2$ & 40 & 30 & 60 & 1164627 \\ \hline
        Aspartic Acid & C$_4$H$_7$NO$_4$ & 70 & 52 & 104 & 10543213 \\ \hline
        Glutamic Acid & C$_5$H$_9$NO$_4$ & 78 & 59 & 118 & 17208382 \\ \hline
        Proline & C$_5$H$_9$NO$_2$ & 62 & 49 & 98 & 8368092 \\ \hline
        Alanine & C$_3$H$_7$NO$_2$ & 48 & 37 & 74 & 2725840 \\ \hline
        Hydrogen & H$_2$ & 2 & 2 & 4 & 15 \\ \hline
        Water & H$_2$O & 10 & 7 & 14 & 1086 \\ \hline
        Carboxy Group & COOH & 23 & 16 & 32 & 54229 \\ \hline
        Amino Group & NH$_2$ & 9 & 7 & 14 & 1086 \\ \hline
        Methylidyne & CH & 7 & 6 & 12 & 631 \\ \hline
   
        R\_His & C$_4$H$_5$N$_2$ & 43 & 34 & 70 & 1978718 \\ \hline
        R\_Leu & C$_4$H$_9$ & 33 & 29 & 58 & 520540 \\ \hline
        R\_Ile & C$_4$H$_9$ & 33 & 29 & 14 & 520540 \\ \hline
        R\_Lys & C$_4$H$_{10}$N & 41 & 35 & 70 & 2197466 \\ \hline
        R\_Met & C$_3$H$_6$S & 40 & 30 & 60 & 506627 \\ \hline
        R\_Phe & C$_7$H$_7$ & 49 & 42 & 84 & 3722223 \\ \hline
        R\_Thr & C$_2$H$_4$O & 24 & 19 & 38 & 49606 \\ \hline
        R\_Trp & C$_9$H$_8$N & 69 & 58 & 116 & 14864603 \\ \hline
        R\_Val & C$_3$H$_7$ & 25 & 22 & 44 & 341819 \\ \hline
        R\_Arg & C$_4$H$_9$N$_3$ & 54 & 44 & 88 & 5411505 \\ \hline
        R\_Cys & CH$_3$S & 25 & 17 & 34 & 100148 \\ \hline
        R\_Gln & C$_3$H$_6$NO & 39 & 31 & 62 & 816630 \\ \hline
        R\_Asn & C$_2$H$_4$NO & 31 & 24 & 48 & 288581 \\ \hline
        R\_Tyr & C$_7$H$_7$O & 57 & 47 & 94 & 4268254 \\ \hline
        R\_Ser & CH$_3$O & 17 & 13 & 26 & 41068 \\ \hline
        R\_Gly & H & 1 & 1 & 2 & 4 \\ \hline
        R\_Asp & C$_2$H$_3$O$_2$ & 31 & 23 & 46 & 375266 \\ \hline
        R\_Glu & C$_3$H$_5$O$_2$ & 39 & 30 & 60 & 1161463 \\ \hline
        R\_Pro & C$_3$H$_6$ & 24 & 22 & 42 & 73108 \\ \hline
        R\_Ala & CH$_3$ & 9 & 8 & 16 & 1977 \\ \hline
    \end{tabular}
\end{table*}
\end{document}